\documentclass{amsart}

\usepackage[dvips]{graphicx}
\usepackage[T1]{fontenc}
\usepackage[utf8]{inputenc}

\newcommand{\rmi}{\textrm i}

\usepackage{amssymb,latexsym,tensor}
\usepackage{eqnarray,amsmath}
\usepackage{mathtools}
\usepackage{hyperref}
\usepackage{url}

\newtheorem{theorem}{Theorem}[section]
\newtheorem{definition}{Definition}[section]

\newtheorem{remark}{Remark}[section]

\numberwithin{equation}{section}
\newcommand{\defeq}{\vcentcolon=}

\author[V. Pasic]{Vedad Pasic}
\author[E. Barakovic]{Elvis Barakovic}
\address{
Department of Mathematics \newline \indent
University of Tuzla \newline \indent
Univerzitetska 4  \newline \indent
Bosnia and Herzegovina}
\email{vedad.pasic@untz.ba \ and \ elvis.barakovic@untz.ba}

\title[Torsion waves in Yang--Mielke gravity]{Torsion Wave Solutions in Yang--Mielke Theory of Gravity}


\begin{document}
\begin{abstract}
 The approach of metric-affine gravity initially distinguishes it from Einstein's general relativity. Using an independent affine connection produces a theory with 10+64 unknowns. We write down the Yang--Mills action for the affine connection and produce the Yang--Mills equation and the so called complementary Yang--Mills equation by independently varying with respect to the connection and the metric respectively. We call this theory the Yang--Mielke theory of gravity. We construct explicit spacetimes with pp-metric and purely axial torsion and show that they represent a solution of Yang--Mills theory. Finally we compare these spacetimes to existing solutions of metric-affine gravity and present future research possibilities.
\end{abstract}
\maketitle

\section{Introduction}
\noindent Einstein's geometric theory of gravity can be summarised, paraphrasing Wheeler \cite{wheeler1990journey}, as follows: spacetime tells matter how to move; matter tells spacetime how to curve. In order to understand this, we have to understand the following:
\begin{itemize}
\item the motion of particles which are so small that their effect on the gravitational field they move in is negligible;
\item the nature of matter as a source for gravity;
\item Einstein's equation, which shows how this matter source is related to the curvature of spacetime.
\end{itemize}
Einstein's equation is at the centre of general relativity.
It gives us a formulation of the relationship between spacetime geometry and
the properties of matter. This equation is formulated using \emph{Riemannian geometry}, in which the geometric properties of spacetime are described by the metric. The vacuum Einstein's equation
$\displaystyle
Ric_{\alpha\beta}-\frac12\mathcal{R}g_{\alpha\beta}=0
$
is obtained by varying the Einstein--Hilbert action
$$\displaystyle
\frac{c^4}{16\pi G} \int \mathcal{R},
$$
with respect to the metric $g$. Here $\mathcal{R}$ is the scalar curvature (\ref{definitionScalar}),
$Ric$ the Ricci curvature \eqref{definitionRicci}, $g$ is the metric, $c$ is the speed of light and
$G$ is the gravitational constant, the recommended numerical value of which is $6.673\ 84(80) \times 10^{-11} \mathrm{m}^3 \mathrm{kg}^{-1}\mathrm{s}^{-2}$, with relative standard uncertainty $1.2 \times 10^{-4}$, see \cite{mohr2012CODATA}.
The full field equation is then obtained by adding the matter Lagrangian to the Einstein--Hilbert
action, which gives us Einstein's equation in tensor form
\begin{equation}\label{einsteins equation} \nonumber
Ric_{\mu\nu} - \frac{1}{2} \mathcal{R} g_{\mu\nu}  = \frac{8\pi G}{c^4} \ T_{\mu\nu},
\end{equation} where $T$ is the stress energy tensor that arises from the matter Lagrangian, see e.g.
\cite{landau1975classical}.
The simplest solution of this equation is the Minkowski spacetime from special relativity.

Two problems with general relativity arose quite quickly after its initial introduction in 1915. Einstein considered that what are recognised locally as inertial properties of matter must be
determined by the properties of the rest of the universe. Einstein's efforts to discover to what extent general relativity manages to do this founded the modern study of cosmology. The second problem of general relativity was that electromagnetism is not included in the theory. As Einstein said in \cite{einstein1967meaning}:
``A theory in which the gravitational field and the electromagnetic field do not enter
as logically distinct structures would be much preferable''.
He expected much more from general relativity than the merging of gravitation and electromagnetism at the macroscopic level and he thought that the theory should explain the existence of elementary particles and should provide a treatment for nuclear forces,  see \cite{einstein1967meaning}. Einstein spent
most of the second part of his life trying to achieve this, unfortunately with no real success and only after his death did the subject again become `fashionable'.

A number of recent developments in physics have evoked the possibility that the treatment of spacetime might involve more than just using the Riemannian spacetime of Einstein's general relativity,  like our failure to quantize gravity, the description of hadron or nuclear matter in terms of extended structures, the accelerating universe, the study of the early universe, etc, see Hehl et al. \cite{hehl1995metric},

The smallest departure from a Riemannian spacetime of Einstein's general
relativity consists of admitting \emph{torsion} (\ref{definitiontorsion}),
arriving at a Riemann--Cartan spacetime, and furthermore, possible \emph{nonmetricity}
(\ref{nonmetricity}), resulting in a \emph{metric-affine} spacetime.
Metric-affine gravity is a natural generalisation of Einstein's
general relativity and it was propagated by Einstein himself for some time. In metric-affine gravity we consider spacetime to be a
connected real 4--manifold $M$ equipped with a Lorentzian metric $g$ and an affine connection
$\Gamma$ and the characterisation of spacetime by an \emph{independent}
linear connection $\Gamma$ immediately distinguishes metric-affine gravity from general relativity.
The connection incorporates the inertial properties of spacetime and it
can be viewed, according to Hermann Weyl \cite{weyl1919neue},
as the guidance field of spacetime, while the metric describes the structure of spacetime with
respect to its spacio-temporal distance relations.

As stated by Hehl et al. in \cite{hehl1995metric}, in general relativity
the linear connection of its Riemannian spacetime is metric-compatible
and symmetric and the symmetry of the Levi--Civita connection translates into the closure of
infinitesimally small parallelograms and the transition from the flat gravity-free Minkowski
spacetime to the Riemannian spacetime in Einstein's theory can locally be
understood as a deformation process.
The lifting of the constraints of metric-compatibility and symmetry
yields nonmetricity and torsion, respectively.
The continuum under consideration is thereby assumed to have
a non-trivial microstructure. The geometrical concepts
of nonmetricity and torsion arise in the three-dimensional continuum theory of lattice
defects, however, see \cite{kroner1981continuum,kroner1986continuized}. For a
comprehensive review of metric-affine gravity, see \cite{hehl1995metric} as well as \cite{blagojevic2013gauge}.

This paper has the following structure. In Section \ref{SectionYangMielke} we introduce the Yang--Mielke theory or gravity. In Section \ref{ppSection} we define our new spacetimes, where in subsection \ref{classicSection} we recall the known properties classical pp-waves and in subsection \ref{generalisedSection} we generalise the notion of a pp-wave to spacetimes with purely axial torsion. Finally, in Section \ref{discussionSection} we compare our solutions with existing results and consider future research possibilities. The appendix provides the notation we use throughout the paper.

\section{Yang--Mielke Theory of Gravity}\label{SectionYangMielke}
We consider spacetime to be a connected real 4-manifold $M$ equipped with a Lorentzian metric $g$
and an affine connection $\Gamma$. The unknowns of our theory are the 10 independent
components of the metric tensor $g_{\mu\nu}$ and the 64 connection coefficients ${\Gamma^\lambda}_{\mu\nu}$.
In \emph{quadratic} metric-affine gravity we define the action as
\begin{equation}\label{action}
S\defeq\int q(R)
\end{equation}
where $q$ is an $O(1, 3)$--invariant quadratic form on curvature $R$. Independent variation of (\ref{action}) with respect to the metric $g$ and the connection $\Gamma$ produces Euler--Lagrange equations which we will
write symbolically as
\begin{align}
\label{eulerlagrangemetric}
\partial S/\partial g&=0\\
\label{eulerlagrangeconnection}
\partial S/\partial \Gamma&=0.
\end{align}
The Yang--Mills action
for the affine connection is a special case of (\ref{action}) with
\begin{eqnarray}\label{YMq}
q(R):=\tensor{R}{^\kappa_{\lambda\mu\nu}}\tensor{R}{^\lambda_{\kappa}^{\mu\nu}}.
\end{eqnarray} The objective of this paper is to study the system of equations (\ref{eulerlagrangemetric}), (\ref{eulerlagrangeconnection}) in the special case (\ref{YMq}).
The origins of this theory lie in the works of \' Elie Cartan, Arthur Eddington, Albert Einstein, Erwin Schr\" odinger, Tullio Levi-Civita and Hermann Weyl. The motivation for choosing a model of gravity which is purely quadratic
in curvature is explained in detail in Section 1 of \cite{vassiliev2005quadratic} and Section 1 of \cite{pasic2014pp}.
The results in this paper strongly rely on the work of C. N. Yang \cite{yang1974integral} and E. W. Mielke \cite{mielke1981pseudoparticle} who respectively showed that Einstein spaces satisfy equations (\ref{eulerlagrangeconnection}) and (\ref{eulerlagrangemetric}). We therefore refer to the special case (\ref{YMq}) of the field equations (\ref{eulerlagrangeconnection}) and (\ref{eulerlagrangemetric}) as the \emph{Yang--Mielke theory of gravity}. There are many works devoted to the study of the system (\ref{eulerlagrangemetric}), (\ref{eulerlagrangeconnection}) in the special case (\ref{YMq}) and one can get an idea of the historical development of the Yang--Mielke theory of gravity from the references stated in \cite{pasic2014pp}. Further aspects of the history of this theory are also given in \cite{mielke2005duality}. Detailed descriptions of the irreducible pieces of curvature and quadratic forms on curvature can be found in \cite{vassiliev2005quadratic,vassiliev2002pseudoinstantons} and our previous work in this theory can be found in \cite{pasic2014pp,pasic2010new,pasic2014new,pasic2005pp}, where much more information and references on the history and development of metric-affine gravity and known solutions of this theory can be found.

\section{PP-waves With Purely Axial Torsion}\label{ppSection}
We use pp-waves, which are well known spacetimes in general relativity, in order to construct our solutions.
In subsection \ref{classicSection} we first provide a brief reminder on classical pp-waves, following the exposition from \cite{pasic2014pp,pasic2005pp} and then in subsection \ref{generalisedSection} we generalise them to spacetimes with purely axial torsion. See \cite{pasic2014pp,pasic2005pp} for extensive references therein for much more information on pp-waves and pp-wave type solutions of metric-affine gravity in general.
\subsection{Properties of classical pp-waves}\label{classicSection}
We define a \emph{pp-wave} as a Riemannian spacetime which admits a nonvanishing parallel spinor field. It is now a well known fact that pp-waves are solutions of the system of equations (\ref{eulerlagrangemetric}), (\ref{eulerlagrangeconnection}), as first shown in \cite{vassiliev2005quadratic}.
We denote the  nonvanishing parallel spinor field appearing in the definition
of pp-waves by $\displaystyle \chi=\chi^a$ and we assume this spinor field to be \emph{fixed}.
Put
$$
l^\alpha:=\sigma^\alpha{}_{a\dot
b}\,\chi^a\bar\chi^{\dot b}
$$
where $\sigma^\alpha$ are Pauli matrices, where we use the spinor formalism introduced in \cite{pasic2014pp,pasic2005pp}. Then
$l$ is a nonvanishing parallel real null vector field. We define the real scalar function, which we call the \emph{phase}, as $$ \varphi:M\to\mathbb{R},\ \varphi(x):=\int l\cdot
d x\,.$$
Put
$ \displaystyle F_{\alpha\beta}:=\sigma_{\alpha\beta
ab}\,\chi^a\chi^b
$
where the $\sigma_{\alpha\beta}$ are `second order Pauli matrices'
$\displaystyle \sigma_{\alpha\beta ac}:=\frac12
\bigl( \sigma_{\alpha a\dot b}\epsilon^{\dot b\dot d}\sigma_{\beta
c\dot d} - \sigma_{\beta a\dot b}\epsilon^{\dot b\dot
d}\sigma_{\alpha c\dot d} \bigr)\,$, see \cite{pasic2014pp,pasic2005pp}.
$F$ can be expressed as
$F=l\wedge m
$
where $m$ is a complex vector field satisfying
\begin{equation}\label{mil}
m_\alpha m^\alpha=l_\alpha m^\alpha=l_\alpha\overline{m}^\alpha=0, m_\alpha\overline{m}^\alpha=-2.
\end{equation}
Note that it is known that pp-waves can also be defined as a
Riemannian spacetime whose metric can be written locally in the form
\begin{equation}
\label{metric of a pp-wave} d s^2= 2d x^0d x^3-(d x^1)^2-(d
x^2)^2 +f(x^1,x^2,x^3)(d x^3)^2
\end{equation}
in some local coordinates $(x^0,x^1,x^2,x^3)$.
The corresponding curvature
tensor $R$ is linear in $f$, i.e.
\begin{equation}\label{curvatureExplicit}
R_{\alpha\beta\gamma\delta}=
-\frac12(l\wedge\partial)_{\alpha\beta}\,(l\wedge\partial)_{\gamma\delta}f,
\end{equation}
where
$(l\wedge\partial)_{\alpha\beta}:=l_\alpha\partial_\beta-\partial_\alpha
l_\beta$.
The choice of local coordinates in which the pp-metric assumes the form
(\ref{metric of a pp-wave}) is not unique. We will restrict our
choice to those coordinates in which
\begin{equation}
\label{explicit l and a} \chi^a=(1,0), \ \ l^\mu=(1,0,0,0),
\ \ m^\mu=(0,1,\mp\rmi,0).
\end{equation}
With this choice of local coordinates (\ref{metric of a pp-wave}), (\ref{explicit l and a}), the phase function becomes
$\varphi(x)=x^3+\mathrm{const}$.
Formula for the curvature of a pp-wave (\ref{curvatureExplicit}) can now be rewritten in invariant form, i.e.
\begin{equation}\label{curvatureRiem}
R=-\frac12(l\wedge\nabla)\otimes(l\wedge\nabla)f,
\end{equation}
where $l\wedge\nabla:=l\otimes\nabla-\nabla\otimes l$.
PP-waves curvature only has two irreducible pieces, i.e. (symmetric) trace--free Ricci and Weyl, see \cite{pasic2014new} for their explicit formulae.

\subsection{Generalising pp-waves to spacetimes with purely axial torsion}\label{generalisedSection}
\begin{definition}
\label{definition of a generalised pp-space} A generalised
pp-wave with purely axial torsion is a metric compatible spacetime with pp-metric and torsion
\begin{equation}
\label{define torsion} T :=  *A
\end{equation}
where $A$ is a real vector field defined by $A = k(\varphi)\ l$, where $k:\mathbb{R}\mapsto \mathbb{R}$ is an arbitrary real function of the phase.
\end{definition}
\begin{remark}
The real vector field $A$ is a plane wave solution of the polarized Maxwell equation $ \displaystyle *d A=\pm\rmi d A$, see equation (16) of \cite{pasic2005pp}. This is not surprising as in our special local coordinates the vector field $A$ is the gradient of a scalar function.
\end{remark}
We list below the main properties of these generalised pp-waves. Note that here and
further on we denote by $\{\!\nabla\!\}$ the covariant derivative
with respect to the Levi-Civita connection which should not be
confused with the full covariant derivative $\nabla$ incorporating
torsion.
Using our special local coordinates (\ref{metric of a pp-wave}),
(\ref{explicit l and a}), we can express torsion as
\begin{equation}\label{torsioNExplicit}
T = \mp \frac{i}{2} k(x^3) \ l\wedge m \wedge \overline{m}.
\end{equation} Note that the $\mp$ sign is chosen to correspond to the sign in (\ref{explicit l and a}).
Torsion (\ref{define torsion}) is clearly purely axial, as $T_{\kappa\mu\nu} = k(\varphi) l^\alpha \varepsilon_{\alpha\kappa\mu\nu}$, which is just the totally antisymmetric part of torsion.

\begin{remark}
Our torsion completely corresponds to Singh's axial torsion from \cite{singh1990axial,singh1990null}. Put $m = -\frac12 k(x^3)$ in formula (16) of \cite{singh1990axial} or put $n=0, m = -\frac12 k(x^3)$ in formula (20) of \cite{singh1990null}.
\end{remark}
\begin{remark}\label{RemarkCompatible}
The connection of a generalised pp-wave with purely axial torsion is clearly metric compatible.
Since $\Gamma^{\kappa}{}_{\mu \nu} = \left\{{{\lambda}\atop{\mu\nu}}\right\} + K^{\kappa}{}_{\mu \nu} = \left\{{{\lambda}\atop{\mu\nu}}\right\} + \frac12 T^{\kappa}{}_{\mu \nu}$, we get that
\[
\nabla_\mu g_{\alpha\beta} = \{\!\nabla\!\}_\mu g_{\alpha\beta} - K^\eta{}_{\mu\alpha}g_{\eta\beta}
- K^\eta{}_{\mu\beta}g_{\alpha\eta} 
\] and $\{\!\nabla\!\}_\mu g_{\alpha\beta} = 0$ as classical pp-waves are metric compatible. However, since our torsion is purely axial, we get that
$
\nabla_\mu g_{\alpha\beta}  =  K_{\alpha\mu\beta} -  K_{\alpha\mu\beta} = 0,
$ i.e. we have metric-compatibility.
\end{remark}
\noindent The curvature of a generalised pp-wave is
\begin{align}
R=&-\frac12(l\wedge\{\!\nabla\!\})\otimes(l\wedge\{\!\nabla\!\})f \label{cuvatureFormula} \\
&+\frac14k^2\ \mathrm{Re} \left( \,(l\wedge m)\otimes(l\wedge \overline{m})\right)
\mp\frac12k' \ \mathrm{Im} \left( \, (l\wedge m)\otimes(l\wedge \overline{m})\right). \nonumber
\end{align}
This can be equivalently written down as
\begin{equation}
R_{\alpha\beta\gamma\delta} =-\frac12(l\wedge\partial)_{\alpha\beta}\,(l\wedge\partial)_{\gamma\delta}f
+ \sum_{i,j=1}^2 r_{ij}(l\wedge m_i)_{\alpha\beta}(l\wedge m_j)_{\gamma\delta}, \label{explicitCurvature}
\end{equation} where $ \displaystyle r_{11}=r_{22} = \frac14 k^2, \  r_{12}=-r_{21}= \pm \frac12 k'$.
It is a highly nontrivial fact that the torsion generated curvature i.e. ${R_T}^\kappa{}_{\lambda\mu\nu} = \partial_\mu{K^\kappa}_{\nu\lambda}
-\partial_\nu{K^\kappa}_{\mu\lambda}
+{K^\kappa}_{\mu\eta}{K^\eta}_{\nu\lambda}
-{K^\kappa}_{\nu\eta}{K^\eta}_{\mu\lambda} $, which is equal to
\begin{equation}
{R_T} =
\frac{k^2}4\!\mathrm{Re} \left(\!(l\wedge m)\!\otimes\!(l\wedge \overline{m})\!\right)
\mp\frac{k'}2\!\mathrm{Im} \left(\!(l\wedge m)\!\otimes\!(l\wedge \overline{m})\!\right) \label{torsionCurv}
\end{equation}  and the Riemannian curvature simply add up to produce formula (\ref{cuvatureFormula}).
We know that the Riemannian part of curvature has two irreducible pieces of curvature, namely Weyl and (symmetric) trace-free Ricci. It turns out that the torsion also generates Ricci curvature and it reads
\[
Ric = \frac12 \left(f_{11} + f_{22}-k^2\right) l\otimes l,
\] where $f_{\alpha\beta}:=\partial_\alpha \partial_\beta f$. Scalar curvature is then clearly zero by the properties of $l$.
\begin{remark}
The Ricci curvature is zero if Poisson's equation
$\displaystyle
f_{11} + f_{22} = k^2
$ is satisfied.
\end{remark}
\begin{remark}
Ricci is parallel if
$ \displaystyle
f_{11}+f_{22}=k^2 + C,
$ in which case
$
Ric = \Lambda \ l\otimes l,\,$ for some constant $\Lambda$.
\end{remark}
\section{New Solutions of Yang--Mielke Theory of Gravity}
In this section we aim to use the spacetimes introduced in Section \ref{ppSection} in order to construct new solutions of the Yang--Mielke theory of gravity. The main result of this paper is the following
\begin{theorem}\label{mainthm}
Generalised pp-waves with purely axial torsion with parallel $\{\!Ric\!\}$ are solutions of (\ref{eulerlagrangemetric}), (\ref{eulerlagrangeconnection}) in the special case (\ref{YMq}).
\end{theorem}
\begin{remark}
Note that by $\{\!Ric\!\}$ we denote the Ricci curvature generated by the Levi-Civita connection only. The condition $\{\!\nabla\!\} \{\!Ric\!\} = 0 $ implies that $f_{11}+f_{22}=C$. Note that the result also holds if $Ric$ is assumed to be parallel.
\end{remark}
\begin{remark}
In the special case (\ref{YMq}), we call equation (\ref{eulerlagrangeconnection}) the \emph{Yang--Mills equation for the affine connection}, i.e.
\begin{equation}\label{YM1}
\partial_\nu R^{\mu\nu} + [\Gamma_\nu,R^{\mu\nu}] = 0,
\end{equation}
where $[\Gamma_\nu,R^{\mu\nu}]^\kappa{}_\lambda=\Gamma^\kappa{}_{\nu\eta} R^\eta{}_\lambda{}^{\mu\nu}
-\Gamma^\eta{}_{\nu\lambda}R^\kappa{}_\eta{}^{\mu\nu}$.
We call the equation (\ref{eulerlagrangemetric}) in the special case (\ref{YMq}) the \emph{complementary Yang--Mills equation}, i.e.
\begin{equation}\label{YM2}
H-\frac{1}{4}(\textrm{tr }H)g = 0,
\end{equation} where $H=H_{\nu}^{\ \rho} :=R^\kappa_{\ \lambda\mu\nu}R^{\lambda}{}_{\kappa}{}^{\mu\rho}$.
Equivalently, equation (\ref{YM2}) can written down as
\[
R^{\kappa}{}_{\lambda\nu}{}^\alpha R^\lambda{}_{\kappa}{}^{\nu\beta}
-\frac14 g^{\alpha\beta}{R^\kappa{}_{\lambda\mu\nu}R^{\lambda}{}_\kappa{}^{\mu\nu}} = 0.
\] See \cite{pasic2009new} for the explicit derivations of the equations (\ref{YM1}) and (\ref{YM2}).
\end{remark}
\begin{proof}[{\bf Proof} of Theorem \ref{mainthm}]
Since we know, see e.g. \cite{pasic2014pp,pasic2005pp,vassiliev2004quadratic}, that classical pp-waves of parallel Ricci curvature are solutions of (\ref{eulerlagrangemetric}), (\ref{eulerlagrangeconnection}) in the special case (\ref{YMq}), it is enough to show the result for the torsion generated part of curvature (\ref{torsionCurv}). In proving that generalised pp-waves solve the equations (\ref{YM1}) and (\ref{YM2}), we will use equations (\ref{mil}), special local coordinates (\ref{metric of a pp-wave}),
(\ref{explicit l and a}), as well as the formulae for curvature (\ref{cuvatureFormula}), (\ref{explicitCurvature}), torsion (\ref{torsioNExplicit}) and torsion generated curvature (\ref{torsionCurv}).
To show that equation (\ref{YM1}) is satisfied, we only need to show that
\[
\partial_\nu {R_T}^{\mu\nu} + [K_\nu,{R_T}^{\mu\nu}] = 0,
\] since the curvature is the sum of the Riemannian curvature (\ref{curvatureRiem}) and the torsion generated curvature (\ref{torsionCurv}), the connection the sum of the Christoffel symbol and contortion and since classical pp-waves solve the Yang--Mills equation. Now, since  $F=l\wedge m$, in special local coordinates (\ref{metric of a pp-wave}), (\ref{explicit l and a}), $F$ takes the form
\[
F^{\alpha\beta}=\left(\begin{array}{cccc} 0 & 1 & \mp i & 0 \\ -1 & 0 & 0 & 0 \\ \pm i & 0 & 0 & 0 \\ 0 & 0 & 0 & 0 \\\end{array}\right)
\] and since the function $k(\varphi)$ is the function of $x^3$ in special local coordinates, using the formula for curvature (\ref{torsionCurv}) we directly get that $\partial_\nu {R_T}^{\mu\nu} = 0$.

Using the explicit formula for torsion (\ref{torsioNExplicit}), the fact that it is purely axial, which implies that $T = 2K$, the explicit formula for torsion generated curvature (\ref{torsionCurv}) and special local coordinates (\ref{metric of a pp-wave}), (\ref{explicit l and a}), we get that the only nonzero term of $\Gamma^\kappa{}_{\nu\eta} R^\eta{}_\lambda{}^{\mu\nu}$ is for $\kappa = 0, \lambda = 3, \mu = 0$, i.e. $-\frac12 k\cdot k'$. However, the only nonzero term of $\Gamma^\eta{}_{\nu\lambda}R^\kappa{}_\eta{}^{\mu\nu}$ is also when $\kappa = 0, \lambda = 3, \mu = 0$, i.e. $-\frac12 k\cdot k' $, so these two terms cancel out. Hence, the equation (\ref{YM1}) is satisfied.

Checking that all the terms in equation (\ref{YM2}) are zero is a straightforward exercise, using the formulae for curvature (\ref{cuvatureFormula}), (\ref{explicitCurvature}), special local coordinates (\ref{metric of a pp-wave}), (\ref{explicit l and a}) and equation (\ref{mil}).
\end{proof}

\section{Discussion and Comparison With Existing Solutions}\label{discussionSection}
Vassiliev \cite{vassiliev2005quadratic} proved the uniqueness of Riemannian solutions of the system of equations (\ref{eulerlagrangemetric}), (\ref{eulerlagrangeconnection}), so we are left with searching for non-Riemannian solutions, i.e. those incorporating torsion and possible nonmetricity. Vassiliev \cite{vassiliev2005quadratic} presented one non-Riemannian solution of the system (\ref{eulerlagrangemetric}), (\ref{eulerlagrangeconnection}) in the most general case of the purely quadratic action (\ref{action}), which was a \emph{torsion wave} solution with explicitly given torsion. This result was previously presented in \cite{king2001torsion} for the Yang--mills case  (\ref{YMq}) and was first independently obtained by Singh and Griffiths \cite{singh1990new}.
Vassiliev went on to conclude that this torsion wave was a non-Riemannian analogue of a pp-wave, whence came the motivation for generalising the notion of a classical Riemannian pp-wave to spacetimes with torsion in such a way as to incorporate the non-Riemannian torsion-wave
solution into the construction.

Constructing vacuum solutions of quadratic metric-affine gravity in terms of pp-waves is a recent development. Classical pp-waves of parallel Ricci curvature were first shown to be solutions of (\ref{eulerlagrangemetric}), (\ref{eulerlagrangeconnection})  in \cite{vassiliev2005quadratic,vassiliev2004quadratic,vassiliev2003pseudoinstantons}.
There are a number of other publications in which authors suggested
various generalisations of the concept of a classical pp-wave, see \cite{obukhov2004generalized} and extensive references therein. These generalisations were performed within the Riemannian setting and usually involved the incorporation of a constant non-zero scalar curvature. Our
construction in \cite{pasic2014pp,pasic2010new,pasic2014new,pasic2005pp}
generalised the concept of a classical pp-wave in a different
direction - we added torsion while retaining zero scalar curvature. We preserved this approach in the current paper.

An interesting generalisation of the concept of a pp-wave was presented by Obukhov in \cite{obukhov2006plane}. Obukhov was motivated by his previous result \cite{obukhov2004generalized} which deals with the Riemannian case. The ansatz for the metric and the coframe of \cite{obukhov2006plane} is exactly the same as in the Riemannian case. However, the connection extends the Christoffel connection so that torsion and \emph{nonmetricity} (\ref{nonmetricity}) are present and are determined by this extension of the connection. Obukhov's gravitational wave solutions provide a minimal generalisation of the pseudoinstanton (see \cite{vassiliev2002pseudoinstantons}), in the sense that nonmetricity does not vanish and that curvature has two irreducible pieces. We chose not to take this approach, as our connection remains metric-compatible.

In our previous work \cite{pasic2014pp,pasic2010new,pasic2014new,pasic2005pp}, we presented results which were new explicit vacuum solutions of the system of our field equations (\ref{eulerlagrangemetric}), (\ref{eulerlagrangeconnection}) and called these generalised pp-waves with torsion. However, it is important to note that the torsion there was \emph{purely tensor}, while the torsion considered in the current paper is \emph{purely axial}. The previous generalisation was done similarly to the approach of this paper, by using the pp-metric and giving an explicit torsion, identical to the torsion-wave obtained by Vassiliev in \cite{vassiliev2005quadratic}. We further explored the properties and characteristics of these generalised pp-waves, showing that they are indeed solutions of the system of our field equations (\ref{eulerlagrangemetric}), (\ref{eulerlagrangeconnection}) in the most general case. Generalised pp-spaces with purely tensor torsion of parallel Ricci curvature appear to admit a sensible physical interpretation, which we explored in detail in \cite{pasic2014pp}, where we gave a comparison with a classical model (namely Einstein--Weyl theory), constructed pp-wave type solutions of this theory and pointed out that generalised pp-waves of parallel Ricci curvature are very similar to pp-wave type solutions of the Einstein--Weyl model. The main difference in using the metric-affine model lies in the fact that Einstein--Maxwell and Einstein--Weyl theories contain the gravitational constant, whereas our model is conformally invariant and the amplitudes of the two curvatures, namely the torsion generated curvature (\ref{torsionCurv}) and the metric generated curvature (\ref{curvatureRiem}) are totally independent.

We aim to be able to do a similar feat using generalised pp-waves with purely axial torsion, i.e. first show that they are solutions of the field equations (\ref{eulerlagrangemetric}), (\ref{eulerlagrangeconnection}) in the most general case, by writing down the field equations explicitly under the assumption of purely axial torsion. We also hope to be able to similarly compare these solutions to the solutions of Einstein--Weyl theory in order to provide a more detailed physical interpretation of these spacetimes.

In \cite{mielke2006cosmological} Mielke and Romero considered axial torsion in the Einstein--Cartan context. We are particularly interested in the section in which the authors recapitulate the minimal coupling of gravity to Dirac fields and where they state that the coupling of axial torsion to the axial current is the only additional term in a prolongation to Riemann--Cartan spacetimes. This will make a significant influence on our reasoning once we attempt to provide a sensible physical interpretation of our solutions.

It is interesting to note that in \cite{guilfoyle1998yang} Guilfoyle and Nolan discuss the initial value problem for the Yang--Mills equations and argue that while the long term existence has been established, it is not expected that Yang's equations have a long term well-posed initial value problem. Of course, the inclusion of torsion can be very problematic in this context.  As stated in the proof of the main result in \cite{guilfoyle1998yang}, Yang's equations constitute a third order quasi-linear system of partial differential equations and fall outside of the wave equations framework used for the initial value problem of general relativity and gauge theory. The authors separate the metric and the connection and propagate them separately by wave equations and state that if the connection and metric are initially compatible, they remain so throughout. It is important to note that Yang was looking for Riemannian (i.e. torsion-free) solutions, so he specialised the equation (\ref{eulerlagrangeconnection}) to the Levi-Civita connection, \emph{after} the variation was carried out. In this case, equation (\ref{eulerlagrangeconnection}) reduces to $\nabla_\lambda Ric_{\kappa\mu} - \nabla_\kappa Ric_{\lambda\mu} = 0$. However, according to \cite{vassiliev2004quadratic}, for a generic 16-parameter action, equation (\ref{eulerlagrangeconnection}) restricted to the Levi-Civita connection becomes $\nabla Ric = 0$. Since our connection is metric compatible as well, see Remark \ref{RemarkCompatible}, and Ricci is assumed to be parallel, we expect this to also be the case for our solutions and this is a matter which we will look at and hopefully answer in the near future.

The two papers of Singh \cite{singh1990axial,singh1990null} were of special interest to us in our task of finding new vacuum solutions of quadratic metric-affine gravity. Singh \cite{singh1990null} presents solutions of the field equations (\ref{eulerlagrangemetric}), (\ref{eulerlagrangeconnection}) for the Yang--Mills case (\ref{YMq}) for a purely \emph{trace} torsion. We are not sure whether these torsion waves can also be used to create new generalised pp-wave solutions of this theory at this time. In \cite{singh1990axial} Singh presents solutions of the field equations (\ref{eulerlagrangemetric}), (\ref{eulerlagrangeconnection}) for the Yang--Mills case (\ref{YMq}) for a purely \emph{axial} torsion and constructed a class of solutions that cannot be obtained using the double duality ansatz, see e.g. \cite{vassiliev2002pseudoinstantons}. Singh used the `spin coefficient technique' from his previous work with Griffiths \cite{griffiths1972some}. Following the reasoning behind the generalised pp-waves of \cite{pasic2005pp} that were shown to be solutions of the field equations (\ref{eulerlagrangemetric}), (\ref{eulerlagrangeconnection}), where we `combined' the pp-metric and the purely tensor torsion waves to obtain a new class of solutions for quadratic metric-affine gravity, we wanted to be able to do the same with purely axial torsion waves. This paper has managed to do just that in the special case (\ref{YMq}), and we hope to be able to obtain the general result. The next obvious step would be to provide a physical interpretation of these new solutions by comparing them to existing classical solutions, like it was done in \cite{pasic2014pp} for purely tensor torsion generalised pp-waves.

\appendix
\section{Notation}
Our notation follows \cite{vassiliev2005quadratic,pasic2014pp,vassiliev2002pseudoinstantons,pasic2010new,pasic2014new,pasic2005pp,king2001torsion}. We denote local coordinates by $x^{\mu}$, where
$\mu = 0, 1, 2, 3$, and write $\partial_{\mu}:=\partial/\partial x^{\mu}.$ We define the covariant derivative of a
vector field as
$\displaystyle
\nabla_\mu v^\lambda:=\partial_\mu v^\lambda+\tensor{\Gamma}{^{\lambda}_{\mu\nu}}v^{\nu}.
$ We define \emph{nonmetricity} $Q$ by
\begin{equation}\label{nonmetricity}
Q_{\mu \alpha\beta}:=\nabla_\mu g_{\alpha\beta}.
\end{equation}
We say that our connection $\Gamma$ is metric compatible if $\nabla g = 0$. The Christoffel symbol
is $\displaystyle \left\{{\lambda\atop {\mu\nu}}\right\}:=
\frac{1}{2}g^{\lambda\kappa}(\partial_\mu g_{\nu\kappa}+\partial_\nu g_{\mu\kappa}-\partial_\kappa g_{\mu\nu}).$ We use the term `parallel' to describe the situation when the covariant derivative of some spinor or tensor field is identically zero. The interval is $ds^2:=g_{\mu\nu}dx^\mu dx^\nu$. We define the action of the Hodge star on a rank $q$ antisymmetric tensor as
\begin{equation*}
(\ast Q)_{\mu_{q+1}\cdots\mu_4}:=(q!)^{-1}\sqrt{|\det g|}Q^{\mu_1\cdots\mu_q}\varepsilon_{\mu_1\cdots\mu_4},
\end{equation*}
where $\varepsilon$ is the totaly antisymmetric quantity, $\varepsilon_{0123}:=+1$.
We define torsion as
\begin{equation}\label{definitiontorsion}
\tensor{T}{^\lambda_{\mu\nu}}:=\tensor{\Gamma}{^\lambda_{\mu\nu}}-\tensor{\Gamma}{^\lambda_{\nu\mu}}.
\end{equation}
We define contortion as
${K^\lambda}_{\mu\nu}:=\frac12
\bigl(
T{}^\lambda{}_{\mu\nu}+T{}_\mu{}^\lambda{}_\nu+T{}_\nu{}^\lambda{}_\mu
\bigr),
$ see \cite{nakahara1998geometry}.
Torsion and contortion are related as
\begin{equation}\label{torsioncontortion}
\tensor{T}{^\lambda_{\mu\nu}}=\tensor{K}{^\lambda_{\mu\nu}}-\tensor{K}{^\lambda_{\nu\mu}}.
\end{equation}
The irreducible pieces of torsion are, see \cite{vassiliev2002pseudoinstantons},
\begin{equation}\label{dijelovi-torzije1}
T^{(1)}=T-T^{(2)}-T^{(3)}, \quad
\tensor{{T^{(2)}}}{_{\lambda\mu\nu}}= g_{\lambda\mu}v_\nu-g_{\lambda\nu}v_\mu, \quad
T^{(3)}=\ast w,
\end{equation}
where $v_\nu=\frac{1}{3}\tensor{T}{^{\lambda}_{\lambda\nu}},\ w_\nu=\frac{1}{6}\sqrt{|\det g|}\tensor{T}{^{\kappa\lambda\mu}}
\tensor{\varepsilon}{_{\kappa\lambda\mu\nu}}$.
The irreducible pieces $T^{(1)},T^{(2)}$ and $T^{(3)}$ are called \emph{tensor torsion}, \emph{trace torsion}, and \emph{axial torsion}
respectively.
The irreducible decomposition of contortion is:
\begin{equation}
K^{(1)}=K-K^{(2)}-K^{(3)},  \quad
\tensor{{K^{(2)}}}{_{\lambda\mu\nu}}=g_{\lambda\mu}v_\nu-g_{\nu\mu}v_\lambda, \quad K^{(3)}=\frac{1}{2}\ast w, \label{contortion1}
\end{equation}
where
$\displaystyle
v_\nu=\frac{1}{3}\tensor{K}{^{\lambda}_{\lambda\nu}},\ \ w_\nu=\frac{1}{3}\sqrt{|\det g|}\tensor{K}{^{\kappa\lambda\mu}}\tensor{\varepsilon}{_{\kappa\lambda\mu\nu}}.
$
The irreducible pieces of torsion (\ref{dijelovi-torzije1}) and contortion (\ref{contortion1}) are related as
$ \displaystyle
\tensor{{T^{(i)}}}{_{\kappa\lambda\mu}}=\tensor{{K^{(i)}}}{_{\lambda\kappa\mu}}\ (i=1,2), \
\tensor{{T^{(3)}}}{_{\kappa\lambda\mu}}=2\tensor{{K^{(3)}}}{_{\kappa\lambda\mu}}.$
We define curvature as
$$\displaystyle
\tensor{R}{^{\kappa}_{\lambda\mu\nu}}:=\partial_\mu\tensor{\Gamma}{^\kappa_{\nu\lambda}}
-\partial_\nu\tensor{\Gamma}{^\kappa_{\mu\lambda}}
+\tensor{\Gamma}{^\kappa_{\mu\eta}}\tensor{\Gamma}{^\eta_{\nu\lambda}}
-\tensor{\Gamma}{^\kappa_{\nu\eta}}\tensor{\Gamma}{^\eta_{\mu\lambda}},
$$
Ricci curvature as
\begin{equation}\label{definitionRicci}
Ric_{\lambda\nu}\defeq\tensor{R}{^{\kappa}_{\lambda\kappa\nu}},
\end{equation} scalar curvature as
\begin{equation}\label{definitionScalar}
\mathcal{R}\defeq\tensor{Ric}{^\kappa_\kappa}
\end{equation} and trace-free Ricci
curvature as $\mathcal{R}ic=Ric-\frac{1}{4}\mathcal{R}g.$ We denote Weyl curvature by $\mathcal{W}$ which is understood as the irreducible piece of curvature defined by the conditions
$$
\tensor{R}{_{\kappa\lambda\mu\nu}}=\tensor{R}{_{\mu\nu\kappa\lambda}}, \ \
\tensor{\varepsilon}{^{\kappa\lambda\mu\nu}}\tensor{R}{_{\kappa\lambda\mu\nu}}=0 \ \textrm{  and  }
\ Ric=0.
$$
Given a scalar function $f:M\rightarrow R$ we write
$$\int f:=\int f \sqrt{\vert \mathrm{det} g \vert}\mathrm{d}x^0\mathrm{d}x^1\mathrm{d}x^2\mathrm{d}x^3,\ \ \mathrm{det} g:=\mathrm{det}(g_{\mu\nu}).$$

\section*{Acknowledgements}
Sincere thanks to Dmitri Vassiliev, Friedrich Hehl and Milutin Blagojevi\' c for helpful advice. We also thank the anonymous reviewers for very useful comments and suggestions.

\bibliographystyle{MyBST}
\bibliography{bibliography}

\label{lastpage}

\end{document}